\documentclass[twocolumn,showpacs,preprintnumbers,amsmath,amssymb]{revtex4}

\usepackage{amsmath}
\usepackage{amsfonts}
\usepackage{amssymb}
\usepackage{graphicx}
\usepackage{subfigure}
\usepackage{graphicx}
\usepackage{dcolumn}
\usepackage{bm}

\begin{document}


\title{Brane Cosmology with the Chameleon Scalar Field in
Bulk }

\author{$^{a,b}$Kh. Saaidi}
 \email{ksaaidi@uok.ac.ir; ksaaidi@phys.ksu.edu}
 \author{$^a$A. Mohammadi}
  \email{abolhassanm@gmail.com}
\affiliation{ %
$^a$Department of Physics, Faculty of Science, University of Kurdistan,  Sanandaj, Iran\\
$^b$Department of Physics, Kansas State University,116 Cardwell Hall, Manhattan, Kansas 66506, USA.\\
}%


\date{\today}

\def\br{\biggr}
\def\bl{\biggl}
\def\Br{\Biggr}
\def\Bl{\Biggl}
\def\be{\begin{equation}}

 \def\ee{\end{equation}}
\def\bea{\begin{eqnarray}}
\def\eea{\end{eqnarray}}
\def\f{\frac}
\def\n{\nonumber}
\def\l{\label}

\begin{abstract}
In this work we investigate the effect of a kind of scalar field,
called chameleon, on the evolution of the Universe. We put this
scalar field in the bulk. { It is displayed that this scalar field
gives us an exponential expansion in early time which may concern inflation. Interaction between  the scalar field and matter brings some complications to our analysis; however, it is shown that by defining an effective potential, we could recover conventional equation at inflation era. After inflation, and entering the Universe in the radiation era, the exponential expansion is omitted. Also, in late time the Universe possesses an accelerated expansion.} In the last
section, the validity of the generalized second law of thermodynamics,
with the assumption of the validity of the first law is considered.
\end{abstract}

\maketitle

\newpage


\section{Introduction}
In the last decades, an increasing amount of data, such as
astrophysical data from type Ia supernovae {\cite{1}}, and cosmic
microwave background radiation, indicate that our Universe
is undergoing an accelerated expansion. Since  ordinary matter
cannot cause this expansion,  another type
of matter has been introduced, called {\it dark energy with a negative pressure},
responsible for this accelerated expansion. Dark energy is one of
the most puzzling aspects of our observed Universe. It seems that
the best and simplest candidate for dark energy is the  cosmological
constant, with an  equation of state $p=-\rho$, but it has some
problems like "fine-tuning". Scalar fields are introduced as
another model for the explanation dark energy. These explanation  treat the  scalar
field as a  dark energy component with a dynamical equation of
state. The dynamical dark energy proposal is often realized by
some scalar field mechanism which suggests that energy formed with
negative pressure is provided by a scalar field evolving down a
proper potential. So far, people have investigated a large class
of scalar field models. A most general model for cosmic
acceleration is a slow-rolling scalar field, called {\it quintessence}
{\cite{2}}. The slow roll means that the scalar field has negative
pressure and then causes  positive accelerating  expansion. Another
attempt to explain dark energy is the  phantom field {\cite{3}}, as
well as modification of  gravitational theory {\cite{4}}. The
equation of state of the  phantom model is displayed by $p=\omega\rho$
where $\omega$ is smaller than $-1$. Another interesting scalar
field model is named "chameleon". This scalar field has been
suggested by Khoury and Weltman {\cite{5}}. In this model a
coupling to the matter is proposed which gives the scalar field a
mass depending on the local density of matter (for a useful reference,see {\cite{6}}); a specific example of a chameleon field arising
from string theory is given in {\cite{7}}. This dependence on the
local density of matter causeas the  chameleon field to have  a suitable
value of mass  and so it  can have a good result in the solar system,
whereas quintessence is not an appropriate model in this scale
because the value of the  mass of the scalar field is small. It is one of the
advantages
of the chameleon scalar field.\\
Another model which has attracted a huge amount of attention is the theory
of extra-dimension where all kinds of matter and their interaction
are confined on a hypersurface (brane), except gravity which can
propagate along the fifth dimension, embedded in a
higher-dimension space-time (bulk). Since dark energy and dark
matter are detected only by its gravitational interaction, it may be
interpreted as the gravitational effects of other branes in the
bulk. Because of this and some other attractor properties, this
model has received much attention recently. The possibility that
our four-dimensional Universe may be embedded in a
higher-dimension bulk space-time is motivated by superstring
theory and M-theory. Higher-dimension models have a long history,
but were revived by the  works of L. Randall and R. Sundrum in 1999
{\cite{8}}. They have introduced two models in order to solve the
hierarchy problem in particle physics; however, after a while these
two models, because of their interesting properties,  attracted
 salient attention in cosmology. In their first model they
consider  two branes of  which our brane has a negative tension (T.
Shiromizu \textit{et al} {\cite{9}} have shown that this model is
unphysical). In their second model they consider a brane with
infinite extra-dimension. In that model our Universe has a
positive tension. The effective four-dimensional gravity in the
brane is modified by extra-dimension {\cite{9,10}}. There is some
correction in generalized Friedmann equation, such as its dependence
on the  quadratic brane energy density that couples directly to the
five-dimensional Planck scale. The classical Friedmann equation
can be recovered in the low energy in late time, when energy
density is much smaller than the brane tension. In the
Randall-Sundrum brane world scenario the bulk contains only a
negative cosmological constant. This needs a fine-tuning between the
cosmological constant of the  bulk with the  tension of a  brane. It may be
more desirable to introduce a model without the  requirement of  fine-tuning. String/M theory suggests that it will also contain a  scalar
field which is  free to propagate through the bulk {\cite{11}},
so it is natural and attractive  to consider the existence of a
scalar field in the bulk. This has been investigated in several
works such as {\cite{11-a,11-b}}. The existence of bulk matter or
scalar field can influence the cosmological evolution on the
brane. The interesting properties of these two models, namely
chameleon scalar field and brane world scenario, have motivated
us to investigate the effects of a chameleon scalar field in bulk
on the evolution on
Universe.\\
The plan of this paper is as follows: In Sec. II, the action
of the brane and bulk is introduced. This action is similar to
{\cite{5}} , as one can see the scalar field is embedded in bulk space-time and this scalar field can interact with matter which is
confined on the brane. { In addition some notes related to gravity localization and basic equations are given. With the help of five-dimensional action and supposing metric fluctuation, the wave equation for transverse-traceless mode of the metric fluctuation is obtained. The wave equation helps us to determine whether the fluctuation mode falls or rises away from the brane, and thus whether or not gravity actually localized the brane. In next part, the junction conditions, basic evolution equations, and scalar field equation of motion are  derived,, and it is shown that the matter density because of interaction with matter is not conserved.  In Sec.
III, the
basic evolution equation on the brane (our Universe) is
acquired. With investigation of the evolution of the  Universe we realize that there is an exponential
accelerated expansion in early times, so we try to have some explanation related to inflation and reheating and consider them in  more detail. By passing time and entering the Universe in the  radiation dominant era, the exponential term is omitted. In late time, the Universe possesses  a positive accelerated expansion.} In Sec. IV, we consider the
validity of the generalized second law of thermodynamics (GSLT) of
this model in an accelerated expanding Universe for the  apparent
horizon and cosmological event horizon. It is indicated that when
matter on the brane is a sort of phantom or quintessence, the
validity of the GSLT is completely confirmed.

\section{General Framework}
To begin our work, we consider the following action

\begin{eqnarray}\label {1}
S & = & \int d^5x \sqrt{-g}\Big\{
\frac{M_5^3}{2}R^{(5)}-\frac{1}{2}(\nabla\phi)^2-V(\phi))
\Big\} \nonumber \\
 & & -\int d^4x \mathcal{L}_m (\psi_m,\tilde{h}_{\mu\nu}).
\end{eqnarray}
The first term describes the five-dimensional gravity in the
presence of a scalar field $\phi$ and the second term describes
the matter of the  brane that is  coupled to the scalar field by

\begin{equation}\label {2}
\tilde{h}_{\mu\nu}=\exp{(\frac{2\beta\phi}{M_p^{3/2}})}
h_{\mu\nu},
\end{equation}
where $\beta$ is a dimensionless coupled constant, and $M_p$ is
the effective four-dimensional Planck mass on the brane. Note that the brane tension is hidden on the last term
action, namely $\mathcal{L}_m$, and tension appears on the brane
energy density in evolution equation (this feature can be seen in
some papers such as {\cite{10,10-a}}). Note that $\mathcal{L}_m$
is a { psedudoscalar density} of weight $1$. { In Eq.(\ref{1}), $g_{\mu\nu}$ is a five-dimensional metric, with signature (-,+,+,+,+), and $h_{\mu\nu}$ is the induced metric of $g_{\mu\nu}$, and denotes four-dimensional metric ofthe  brane. Matter minimally couples to $\tilde{h}_{\mu\nu}$ related to the induced metric $h_{\mu\nu}$ by a conformal transformation.}
One can obtain the scalar field equation  by varying the motion with respect to $\phi$ as

\begin{equation}\label {3}
\nabla ^2\phi=V_{,\phi}(\phi)+ \frac{1}{\sqrt{-h}}
\frac{2\beta}{M_p^{3/2}} \frac{\partial\mathcal{L}_m }{\partial
\tilde{h}_{\mu\nu}}\tilde{h}_{\mu\nu} \delta(y),
\end{equation}
{ where $V(\phi)$ is a potential of the chameleon scalar field,
which is almost flat.}  More information about the potential and
its feature can be found in {\cite{5,6}}. Varying with respect to
the metric $g_{\mu\nu}$, the Einstein equation is obtained as

\begin{equation}\label {4}
^{(5)}G^{\mu\nu}=\kappa_5^2 \Big\{ T^{(\phi)\mu\nu}+T^{(b)\mu\nu}
\Big\},
\end{equation}
where $T^{(\phi)\mu\nu}$ stands for the scalar field of total
energy-momentum tensor $$T^{(\phi)\mu\nu}=\nabla ^{\mu} \phi
\nabla ^{\nu} \phi - g^{\mu\nu}
(\frac{1}{2}(\nabla\phi)^2+V(\phi)).$$ With attention to the
dimension of the component of the energy-momentum tensor, we realize
that the dimension of the scalar field is no longer $M$, rather
it is $M^\frac{3}{2}$. Because of  this fact, we select
$M_p^{\frac{3}{2}}$ in the exponential of Eq.(\ref{2}), to make
it dimensionless. Also the dimension of potential is $M^5$. Since
we put scalar field in five-dimensional space-time, all of these
results are acquired. $T^{(b)\mu\nu}$ stands for the brane energy-momentum tensor part of the  total energy-momentum tensor, which is
$$T^{(b)\mu\nu}=\frac{2}{\sqrt{-h}} \frac{\partial \mathcal{L}_m
}{\partial h_{\mu\nu}} \delta(y),$$ where it can be rewritten as
$$T^{(b)\mu}_{\nu}= \frac{\delta(y)}{b} diag (-\rho_b,p_b,p_b,p_b,0).$$
\par Assuming that types of  matter do not interact with each other, so the
energy-momentum tensor is conserved in the  A frame {\cite{6}}, namely,

\begin{equation}\label {5}
\tilde{D}_\mu \tilde{T}^{(b)\mu\nu}=0,
\end{equation}
where $\tilde{D}$ indicates the covariant derivative in
four-dimensional  space-time, and $\tilde{T}^{(b)\mu\nu}$ is
described by

\begin{equation}\label {6}
\tilde{T}^{(b)\mu\nu}=\frac{2}{\sqrt{-\tilde{h}}} \frac{\partial
\mathcal{L}_m }{\partial \tilde{h}_{\mu\nu}} \delta(y).
\end{equation}
This equation gives us the conservation relation in the  A frame,
$$\dot{\tilde{\rho}}+3\tilde{H}(\tilde{\rho}+\tilde{p})=0.$$ {  We
suppose that the fluid in the brane be a perfect fluid with
$\tilde{p}=\omega \tilde{\rho}$, so}

\begin{equation}\label {7}
\tilde{\rho}=\tilde{\rho}_0 \tilde{a}^{-3(1+\omega)}
\end{equation}
where $\tilde{\rho}_0$ is a constant. In the  above equations $y$
displays the coordinate of the  fifth dimension and $\delta(y)$
explains that our matter is confined to a four-dimensional
hypersurface, namely our Universe.

\subsection{Gravity Localization on Branes}
{ Although this paper is focused on the evolution of the  Universe by assuming a scalar field on bulk which is coupled with matter, we
will motivate a class of localized graviton on the brane.
In fact,  as we will show below, this motivation can be partly justified.
The five-dimensional action of the model in the bulk (\ref{1})  can be   expressed as
\begin{equation}\l{8}
S=\int d^5x \sqrt{-g} \Big( \frac{M_5^3}{2} R -\frac{1}{2} g^{MN} \partial_M \phi \partial_N \phi - V(\phi) \Big)
\end{equation}
where there is a scalar field $\phi$ with a potential $V(\phi)$ and the  five-dimensional Ricci scalar, $^5R$. Here the Latin index runs from 0...4 and the Greek index runs from 0...3. The space-time metric is supposed as
\begin{equation}\l{9}
ds^2=g_{MN}dx^M dx^N
\end{equation}
and one can redefine the above metric as follows
\begin{equation}\l{10}
ds^2=e^{2A(y)}  h_{\mu\nu}(x)dx^{\mu} dx^{\nu}+dy^2
\end{equation}
where $y$ stands for extra-dimension and $e^{2A(y)}$ is the warp factor. Here $h_{\mu\nu}$ is four-dimensional metric. We can have the  following definition for the  metric (\ref{23}):
\begin{equation}\l{11}
ds^2=e^{2A(z)} \Big(h_{\mu\nu}(x)dx^{\mu} dx^{\nu}+dz^2 \Big)
\end{equation}
where a coordinate transformation has been imposed as $dz=e^{-A(y)} dy$. In order to study the localized gravity  on the 3-brane, one should consider the equation of motion for the linearized metric fluctuation (\ref{11}). Let us consider the metric fluctuation $\delta g_{MN}=e^{2A(z)} h_{MN}$. The metric is rearranged as
\begin{equation}\l{12}
ds^2=e^{2A(z)} \Big( (\hat{h}_{\mu\nu}(x)+\bar{h}_{\mu\nu}(x,z))dx^{\mu} dx^{\nu}+dz^2 \Big)
\end{equation}
where the axial gauge $\bar{h}_{5M}=0$ has been imposed on the  metric. In general, because gravity is coupled to the scalar field, fluctuation of the scalar field should be considered at the same time that we study the  fluctuation of the  metric background. However, following \cite{25}, we can only investigate transverse-traceless (TT) modes of the metric fluctuation, namely $\bar{h}^{TT}_{\mu\nu}$. In fact the scalar fluctuation vanished under TT gauges. Therefore the dynamics equation of transverse-traceless modes of the metric fluctuation can be written as {\cite{25}}
\begin{equation}\l{13}
\Big( \partial_z^2 - 3(\partial_z A)\partial_z -  \hat{g}^{\alpha\beta} \nabla_{\alpha} \nabla_{\beta} \Big) \bar{h}^{TT}_{\mu\nu} (x,z) = 0.
\end{equation}
The equation could have a solution which is given by $\bar{h}^{TT}_{\mu\nu} (x,z) = K_{\mu\nu} e^{ipx}$ (where $K_{\mu\nu}$ is a constant tensor and $p^2=-m^2$). Finally by using Kaluza-Klein(KK) decomposition, the Schrodinger equation can be achieved as
\begin{equation}\l{14}
\Big( -\partial_z^2 A + V(z) \Big) \psi(z)=m^2 \psi(z)
\end{equation}
where the localizing potential is defined as
\begin{equation}\l{15}
V(z)=\frac{3}{2} \Big[ \partial_z^2 A + \frac{3}{2}(\partial_z A)^2 \Big]
\end{equation}
for more detail refer to {\cite{25}}. According to {\cite{26}}, it is realized that for trapping the massless mode of gravity, the potential $V(z)$ should have a well with a negative minimum inside the brane and satisfy $V(z)>0$ far from the brane, namely for $z\rightarrow \pm \infty$.
By setting $m=0$, the zero mode wave function can be expressed  as $\Psi_0(z)= A_0 \exp(3A(z)/2)$, where $A_0$ is a constant. In order to localize
the four-dimensional gravitation, $\Psi_0 (z)$ has to obey the normalization constraint
$$\int \|\Psi_0(z)\|^2 dz = c,$$
where $c$ is  a finite constant.  It is well-known that the character of  graviton localization  depends on the potential $V_{QM}$ and also depends on the warp factor. In the  braneworld scenario, the four-dimensional effective action is obtained from the five-dimensional action as
\begin{equation}\l{16}
S \sim M_5^3 \int d^5x\sqrt{-g} R_5 \sim M_{p}^2 \int d^4x\sqrt{-h} \hat{R}_4
\end{equation}
where $M_p$  is the four-dimensional Planck scale. The localized zero mode will cause a four-dimensional Newtonian interaction potential. In \cite{22} a de Sitter thick brane type of  action (\ref{9}) is investigated and the authors  have obtained  the gravitational potential between two pointlike masses on  the brane. They find that the  effective potential between two pointlike masses is from the contribution of the zero mode and the continuum KK modes, and is expressed as {\cite{22}}
\begin{equation}\l{17}
U(r)=G_N \frac{M_1 M_2}{r} + \frac{M_1 M_2}{M_5} \int dm \frac{e^{-mr}}{r}\mid \psi_m(0) \mid ^2
\end{equation}
where the first term is a standard Newtonian potential related to the contribution of the  zero mode, and the second term is the correction of the  Newtonian potential related to the contribution of th KK modes.\\
This work is focused on the evolution of the  Universe by assuming a scalar field on bulk which is coupled with matter. The investigation of gravity localization in detail in the model needs more  offers and studying.\\}


\subsection{Equations of Motion on the Brane}
Since we are interested in  study the positive accelerating expansion of the Universe, we introduce a special case of metric instead of (\ref{9}).
So  we continue our work with  a Friedmann-Lema$\hat{\rm i}$tre-Robertson-Walker  FLRW metric as
\begin{equation}\label {18}
ds^2=-n^2(t,y)dt^2+a^2(t,y)\gamma_{ij}dx^idx^j+b^2(t,y)dy^2,
\end{equation}
with a maximally symmetric 3 geometry $\gamma_{ij}$. We assume
that the  brane is embedded on $y=0$, also we take account
$Z_2$ symmetry. It should be mentioned that the metric is
continuous but their first derivative with respect to $y$ is
discontinuous, and their second derivative with respect to $y$
includes the Dirac delta function. Substituting the above metric,
one can obtain the nonvanishing component of the  Einstein tensor in
the following form
\begin{eqnarray}\l{19}
^{(5)}G_{00}&=&-\frac{3}{a^2}\Big\{ -\dot{a}^2+a'^2n^2+a a'' n^2
\Big\},\\
^{(5)}G_{ij}&=&\frac{1}{n^3} \Big\{
2a\ddot{a}n+2\dot{n}a\dot{a}+2n^2n'aa' \nonumber \\
 & &-n\dot{a}^2+a'^2n^3+2a a'' n^3+n^2a^2 n'' \Big\} \delta_{ij},\l{20}\\
^{(5)}G_{05}&=&-\frac{3}{an}\Big\{ \dot{a}'n-n'\dot{a} \Big\},\l{21}\\
^{(5)}G_{55}&=&\frac{3}{a^2n^3}\Big\{
-a\ddot{a}n+\dot{n}a\dot{a}+n^2n'aa'-n\dot{a}^2 \nonumber \\
 & & +a'^2n^3 \Big\}.\l{22}
\end{eqnarray}
Note that in the above equations we take $b(t,y)=1$, and dot
denotes derivative with respect to  time and the prime denotes
the derivative with respect to the fifth coordinate. Since, the second
derivative of metric consists of  the Dirac delta function, according
to {\cite{10}} one can define it as
$$a''=\hat{a}''+[a']\delta(y),$$ where $\hat{a}''$ is the
nondistributional part of the double derivative of $a(t,y)$, and
$[a']$ is the jump in the first derivative across $y=0$, which is
defined by
$$[a']=a'(0^+)-a'(0^-).$$ The junction functions can be obtained
by matching the Dirac delta function in the component of Einstein
tensor with the component of the  brane energy-momentum tensor. From
the $(0,0)$ and $(i,j)$ component of the  field equation we have,
respectively
\begin{eqnarray}\label {23}
\frac{[a']}{a_0}&=&-\frac{\kappa_5^2}{3}\rho_b,\\
\frac{[n']}{n_0}&=&\frac{\kappa_5^2}{3}(2\rho_b+3p_b),\label {24}
\end{eqnarray}
(note that, here, $\rho_b$ and $p_b$, the brane energy density
and pressure respectively, include  the tension of the  brane, namely
$\rho_b=\rho+{\sigma}$ and $p_b=p-{\sigma}$, where $\rho$
and $p$ are matter density and pressure respectively; see
{\cite{10,10-a}}). These equations are the  same as  the junction
relations that { Binetruy , Deffayet, and Longlois}{\cite{10}} have
obtained in their paper. However, we should note here
that the energy density and pressure, namely $\rho_b$ and $p_b$ ,
depend on  the scalar field. We shall explain their relation later.
One can obtain a junction condition for scalar field from its
equation of motion. According to Eq. (\ref{3}), we arrive at
\begin{eqnarray}\label {25}
\frac{\ddot{\phi}}{n^2}-\phi''&+&\left(
\frac{3\dot{a}}{an^2}-\frac{\dot{n}}{n^3}
\right)\dot{\phi}-\left( \frac{n'}{n}+\frac{3a'}{a} \right)\phi'=\nonumber \\
 &-&V_{,\phi}(\phi)-
\frac{2\beta}{{M_p^{3/2}}}\frac{1}{\sqrt{-h}}\frac{\partial\mathcal{L}_m}{\partial
\tilde{h}_{\mu\nu}}\tilde{h}_{\mu\nu} \delta(y).
\end{eqnarray}
Matching the Dirac delta function on  both sides of this relation,
for $y=0$, we have
\begin{equation}\label {26}
[\phi']=\frac{2\beta}{{M_p^{3/2}}}
\frac{1}{\sqrt{-h}}\frac{\partial\mathcal{L}_m}{\partial
\tilde{h}_{\mu\nu}}\tilde{h}_{\mu\nu}.
\end{equation}
The right-hand side of this equation is computed as follows
\begin{eqnarray}\label {27}
\frac{\beta}{{M_p^{3/2}}}
\frac{1}{\sqrt{-h}}\frac{\partial\mathcal{L}_m}{\partial
\tilde{h}_{\mu\nu}}\tilde{h}_{\mu\nu}&=&-\exp{(
\frac{4\beta\phi}{{M_p^{3/2}}} )} \nonumber \\
 & \times & \frac{\beta}{M_p^{\frac{3}{2}}}\frac{-2}{\sqrt{-\tilde{h}}}\frac{\partial\mathcal{L}_m}{\partial
\tilde{h}_{\mu\nu}}\tilde{h}_{\mu\nu} \nonumber \\
 &=&-\exp{(
\frac{4\beta\phi}{{M_p^{3/2}}} )} \frac{\beta}{{M_p^{3/2}}}
\left( \tilde{T}^{\mu\nu}
\tilde{h}_{\mu\nu} \right) \nonumber \\
 &=&\frac{\beta (1-3\omega)}{{M_p^{3/2}}} \tilde{\rho} \exp{(
\frac{4\beta\phi}{{M_p^{3/2}}} )},
\end{eqnarray}
where $\tilde{\rho}$ and $\tilde{p}$ are the energy density and
pressure respectively in the A frame. On the brane, the
energy-momentum tensors in the  Einstein frame and A frame are
related to each other by
\begin{equation}\label {28}
T^{\mu\nu}=\exp(\frac{6\beta\phi}{{M_p^{3/2}}})\tilde{T}^{\mu\nu},
\end{equation}
therefore, $\rho_b$ and $p_b$ are easily expressed in terms of the
component of the  energy momentum of the  A frame and scalar field as
\begin{eqnarray}\label {29}
\rho_b &=& \tilde{\rho}_0 \exp{(
\frac{(1-3\omega)\beta\phi}{{M_p^{3/2}}} )} a_0^{-3(1+\omega)},\\
p_b &=& \omega \tilde{\rho}_0 \exp{(
\frac{(1-3\omega)\beta\phi}{{M_p^{3/2}}} )} a_0^{-3(1+\omega)},\label {30}
\end{eqnarray}
{ whhere $\tilde{\rho}_0$ is a constant that has been introduced in Eq.(\ref{7}) and $a_0$ is a scale factor that is taken on the  brane, namely $y=0$. }
When  the scalar field approaches   zero,  $\rho_b =
\tilde{\rho}$ and $p_b = \tilde{p}$. Also, because of the
presence of $Z_2$ symmetry, as we have assumed before, one can
attain $a'_0$, $n'_0$ and $\phi'_0$ functions from the junction
conditions.
{ From  the $(0,5)$ component of the  field equation we have  $$^{5}G_{05}=\kappa_{5}^{2} T^{(\phi)05}=\kappa_{5}^{2} \dot{\phi} \phi',$$ ($T^{(b)05}=0$). Now, by substituting $\dot{a}'$ and $n'$ from  the junction conditions, (\ref{23}), (\ref{24}), and assuming $Z_{2}$ symmetry, one can obtain the   generalized continuity equation on the brane as}
\begin{equation}\label{31}
\dot{\rho_b}+3H(\rho_b+p_b)=2\dot{\phi_0}\phi'_0.
\end{equation}
As we expected the energy, due to the interaction between matter
and the  scalar field, is not conserved. Note that in all of the relations
in this section we select $n_0=1$, without any loss of generality.
 We recognize that the generalized
continuity equation explicitly confirms the junction condition. From the  $(5,5)$ component of the  field equation, one can
obtain the second order (or generalized) Friedmann equation as
\begin{equation}\label{32}
\frac{\ddot{a}_0}{a_0}+\frac{\dot{a}_0^2}{a_0^2}=-\frac{\kappa^4_5}{36}\rho_b(\rho_b+3p_b)
- \frac{\kappa^2_5}{3}(\frac{\dot{\phi}^2_0}{2}+
\frac{\phi'^2_0}{2} - V(\phi_0)).
\end{equation}
 The first Friedmann equation on the brane is obtained
from $(0,0)$ component of field equation as
\begin{equation}\label{33}
H^2=\left(\frac{\dot{a_0}}{a_0}\right)^2=\frac{\kappa^4_5}{36}\rho_b^2+
\frac{\kappa^2_5}{3}(\frac{\dot{\phi}^2_0}{2}+ \frac{\phi'^2_0}{2}
+ V(\phi_0))+\frac{\hat{a}''_0}{a_0}.
\end{equation}
$\hat{a}''_0$ is the nondistributional part of the  double derivative
of $a(t,y)$ with respect to the fifth coordinate, and the
subscript $0$ means that they are taken in $y=0$. { By ignoring $\frac{\hat{a}''_0}{a_0}$ and using $\rho_b= \rho+ \sigma$, one can obtain
 \begin{equation}\label{34}
H^2=\frac{\kappa^4_5\sigma}{18}\rho{\biggr (}1 + \frac{\rho}{2\sigma}{\biggl )}+\frac{\kappa^4_5}{36}\sigma^2+
\frac{\kappa^2_5}{3}{\biggr(}\frac{\dot{\phi}^2_0}{2}+ \frac{\phi'^2_0}{2}
+ V(\phi_0){\biggl )}.
\end{equation}
It is seen that Eq.(\ref{34}) agrees with the results which were obtained in \cite{8, 8'} for brane world cosmology and is completely different from standard cosmology model, because in standard cosmology $H \propto\sqrt{\rho}$   rather than $ {\rho}$.}
\section{ The  Evolution of the  Universe}
 In this section we want to investigate the behavior of the evolution of the  Universe in early
 and late times. One can rewrite the
generalized Friedmann equation as follows
\begin{equation}\label{35}
\frac{\ddot{a_0}}{a_0}=-\frac{\kappa^4_5}{36}(2+3\omega)\rho_b^2 -
\frac{\kappa^2_5}{3}(\dot{\phi}^2_0+ \phi'^2_0).
\end{equation}
From the junction condition, and with the help of the
function of energy density, we rearrange this relation
as
\begin{eqnarray}\label{36}
\frac{\ddot{a_0}}{a_0}&=&- \frac{\kappa_5^4}{36}(2+3\omega)
\left( \frac{\tilde{\rho}_0}{a_0^{3(1+\omega)}}\right)^2 \nonumber \\
 & & \times \exp{(\frac{2(1-3\omega)\beta\phi}{M_p^{3/2}})} -
\frac{\kappa_5^2}{3}\dot{\phi}^2
\nonumber \\
  & &-\frac{\kappa_5^2}{3} \frac{\beta^2}{4M_p^3}(1-3\omega)^2
  \left( \frac{\tilde{\rho}_0^2}{a_0^{6(1+\omega)}} \right)^2 \nonumber \\
   & & \times  \exp{(\frac{2(1-3\omega)\beta\phi}{M_p^{3/2}})}
\end{eqnarray}
\subsection{ Early time}
{  In early times, namely in the inflation stage, the scalar field is dominant and also it is well-known that the scalar potential energy of the inflation dominates over the kinetic energy. So Eq. (\ref{35}) reduces to
 \begin{eqnarray}\label {37}
\frac{\ddot{a}}{a} &\simeq& - \frac{\kappa_5^4}{3}{\phi'}_0^2,\nonumber\\
&=&  -\frac{\kappa_5^4}{3}\frac{\beta^2 (1-3\omega)^2}{{M_p^{3}}} {\tilde{\rho_0}^2\over a^{6(1+\omega)}}\exp{\biggr \{}{8\beta\phi_0 \over
M_p^{3/2}}{\biggl \}}.
\end{eqnarray}
 The effect of $[\exp{\{4\beta\phi_0/m_p^{3/2}\}}/a^{3(1+\omega)}]^2$  on the  evolution of the  Universe is obvious in this area.  This term  is large and  may explain inflation in very early step of  the Universe evolution. At the end of inflation the evolution of the Universe has a  transition from a de Sitter stage, during which the evolution of the Universe is dominated by the scalar field, to a subsequent
radiation- or matter-dominated Friedmann-Robertson-Walker type cosmological
model. One of the possible approaches to this problem is phenomenological \cite{17, 20, 21}. At this stage the first term of (\ref{35}) is dominated and then we have
\begin{equation}\label {38}
\frac{\ddot{a}}{a} \simeq - \frac{\kappa_5^4}{36}(2+3\omega)
\frac{\tilde{\rho}_0^2}{a_0^{6(1+\omega)}}
\exp{(\frac{2(1-3\omega)\beta\phi}{M_p^{3/2}})}.
\end{equation}
In radiation dominant with $\omega=1/3$, the power of the  exponential function vanishes and we obtain a deceleration phase of expansion for the  Universe.  However, in this stage we want to investigate the inflation and then reheating process of the evolution after inflation.}
\subsubsection{Inflation}
 In the widely accepted inflationary scenario it is assumed that during an initial period
the Universe is dominated by a large, approximately constant potential term $V(\phi)$ of a scalar
field $\phi_0$, known as the inflaton field \cite{16,17}. In our model the energy-momentum tensor of the scalar field
can be written on the brane  in the perfect fluid form, with energy density $\rho_{\phi}$ and pressure $p_{\phi}$ as
\begin{eqnarray}\l{39}
\rho_{\phi_0} &=& {1\over 2} \phi_0^2 + \tilde{V}_{eff}(\phi_0),\\
p_{\phi} &=& {1\over 2} \phi_0^2 - \tilde{V}_{eff}(\phi_0),\l{40}
\end{eqnarray}
where
\be\l{41}
\tilde{V}_{eff}(\phi_0)= V(\phi_0) + {1\over 2} {\phi_0'}^2
\ee
and $\phi_0= \phi(y, t)|_{y=0} = \phi_0(t)$. So in this case Eq. (\ref{34})  reduces to
  \begin{equation}\label{42}
H^2=\frac{\kappa^4_5}{36}\sigma^2+
\frac{\kappa^2_5}{3}\rho_{\phi_0}.
\end{equation}
 Let us now consider the equation of motion of the  scalar field on the
brane.  Equation (\ref{3}) can be rewritten on the brane as
\begin{equation}\label {43}
\ddot{\phi_0}+3H\dot{\phi_0}+\frac{\kappa_5^2}{3}(\rho_b-3p_b)\phi_0'=-V_{,\phi}(\phi_0).
\end{equation}
We have ignored the nondistributional part of $\phi''$. This equation clearly shows the effect of bulk on the equation of motion of $\phi$. We may
introduce an effective potential to describe the dynamic of the
scalar field as
\be\l{44}
V_{eff,\phi}(\phi_0) \equiv V_{,\phi}(\phi_0)+\frac{\kappa_5^2}{3}{\biggr [}\{1-3\omega\}  \rho_b\phi_0'{\biggl ]}.
\ee
  Dependence on energy density of the  effective potential and a  quantity that has  come from bulk is explicit.
 Hence, one can easily write the equation of motion for  the scalar
field on the brane   as
\begin{equation}\label {45}
D^2\phi=-V_{eff,\phi}(\phi_0),
\end{equation}
where $D^2$ is the D'Alembert of the  scalar field in four-dimensional
space-time. It is the  same as the equation
 which is obtained for low energy, four-dimensional  {\cite{5,6}}.    So   one can obtain the mass of the scalar field $\phi$
as
\begin{eqnarray}\l{46}
m^2_{\phi} &=& V_{eff, \phi\phi} , \nonumber\\
&=& V_{,\phi\phi} + {\kappa_5^2\over3}\{1-3\omega \} {\partial \over\partial\phi} {(}  \rho_b\phi_0'{)} .
\end{eqnarray}
Therefore, although bulk is free of  matter, but due to interaction between e matter and the scalar field, the scalar field which propagates through the bulk takes a mass which  is dependent on the brane density energy.  In fact this means that the mass of the  scalar field in the four dimensional effective action is not small and so the correction to the Newton law cannot  be large because of  the propagation of the scalar field in the bulk. We want to obtain a simple  form for $m^2_{\phi}$.
Using (\ref{24}) and  (\ref{25})  we have
\begin{eqnarray}\l{47}
\phi_0' &=& {1\over 2}[\phi_0'], \nonumber\\
&=& \frac{\beta (1-3\omega)}{{M_p^{3/2}}} {\tilde{\rho_0}\over a^{3(1+\omega)}}\exp{\biggr \{}{4\beta\phi_0 \over
M_p^{3/ 2}}{\biggl \}}.
\end{eqnarray}
Using (\ref{29}), (\ref{30}) and (\ref{47}) we have
 \begin{eqnarray}\l{48}
\frac{\kappa_5^2}{3}(\rho_b-3p_b)\phi_0'&=&A(t) \exp{\biggr \{}{\beta(5-3\omega)\phi_0 \over M_p^{3/ 2}}{\biggl \}},
\end{eqnarray}
where $$A(t) = \frac{\beta\kappa_5^2{\tilde{\rho}_0}^2 (1-3\omega)^2}{3a^{6(1+\omega)}M_p^{3/2}}.$$
Substituting  (\ref{48}) in (\ref{44}) and  (\ref{46})  gives
\begin{eqnarray}
V_{eff, \phi}&=&V_{,\phi}(\phi)+A(t) \exp{\biggr \{}{4\beta\phi_0 \over
M_p^{3/ 2}}{\biggl \}}\l{49}, \\
m_{\phi}^{2} &=& V_{,\phi\phi} +{4\beta\over
M_p^{3/ 2}} A(t) \exp{\biggr \{}{4\beta\phi_0 \over
M_p^{3/ 2}}{\biggl \}}.\l{50}
\end{eqnarray}
It is well-known that   the potential energy of the inflation
dominates over the kinetic energy; this means $V_{eff}(\phi_0) \gg \dot{\phi}_0^2/2$. Hence one requires a
 flat potential for the inflation in order to lead
to sufficient inflation. Implying the slow-roll conditions,
$V_{eff}(\phi_0) \gg \dot{\phi}_0^2/2$ and $|\ddot{\phi}_0 | \ll 3 H |\dot{\phi}_0 | $
 Eqs. (\ref{42}) and (\ref{43}) are approximately given as
\begin{eqnarray}
  H^2 &=& {\kappa^2_5 \over 3 } V_{eff}(\phi_0),\l{51}\\
  3H \dot{\phi_0}&=& -V'_{eff}(\phi_0),\l{52}
\end{eqnarray}
the velocity is negative for $V_{eef} >0 $, because the field rolls down the potential towards smaller $\phi_0$ values.

\subsubsection{Reheating after inflation}
{ During a second period of evolution, the potential minimum is approached,
$V_{eff}(\phi_0)$ tends to zero, and the scalar field starts to fluctuate violently around the minimum
value. In fact
at the end of inflation some of the  scalar field energy density needs to be converted to conventional matter to restore hot big bang cosmology usually this process comes by decay of the scalar field (inflaton field). In addition there is the possibility of reheating by gravitational particle production, where the required particles are produced quantum mechanically from  the time varying gravitational field. The method of studying for this kind of reheating depends on the  scalar field equation of motion and the  expansion rate, namely,  Eqs.(\ref{43}) and (\ref{42}) {\cite{30}}. However, in the postinflationary stage, the inflaton
field executes coherent oscillations about the minimum of the potential \cite{34} and the
kinetic term dominates the potential term in the reheating era. Therefore we can expand the effective potential around the minimum point and in the reheating area  we have
\begin{equation}\l{53}
V_{eff}(\varphi) \simeq {1\over 2} m^2_{ \phi_{\rm min}} \varphi^2,
\end{equation}
where we have ignored a constant in (\ref{53}),  $\varphi = \phi_0(t) -\phi_{0_{\rm min}}$ and    the effective potential is minimized at $\phi_{0_{\rm min}}$.
Then the scalar field equation of motion in the  reheating area is
\begin{eqnarray}\l{54}
\ddot{\varphi} + 3H \dot{\varphi} +m^2_{ \phi_{\rm min}}\varphi = 0,
\end{eqnarray}
 According to the   theory of reheating \cite{31, 32}, which  was based on the concept of single-body decays,
 the inflaton field is a collection of scalar
particles each with a finite probability of decaying.
Such decays can be treated by coupling $\phi$
 to other scalar  or fermion fields through
terms in the Lagrangian. We assume the inflaton field is coupled with a matter scalar field as $K \varphi \psi^2$ . Here $\psi$ is a  matter scalar field. Since the homogeneous part of the inflaton is very large
at the end of inflation it behaves  like a classical field.  So  the   inflaton   is treated as a classical external force and acting on the quantum fields $\psi$.
The explicit expression of the decay width of the scalar field can be represented as \cite{33}
\begin{equation}\l{55}
\Gamma = \alpha_{\phi}m_{\phi}\sqrt{1-({T\over m_{\phi}})^2}
\end{equation}
where $\alpha_{\phi}$ and $m_{\phi}$ are the coupling constant and the mass of inflation,  respectively. T is the decaying temperature and it can be related to the matter density, $\rho_m$,  as
\begin{equation}\l{56}
\rho_m = \sigma T^{\gamma \over \gamma -1},
\end{equation}
where $\sigma$ is a constant. For a  radiation dominant, $\rho_m = \pi^2T^4/15$,
$\sigma = \pi^2/15$, and for a   matter-dominated Universe, the relation between $\rho_m$ and $T$ can also
be written down explicitly. The scalar field is negligibly small in the matter-dominated
phase and in the nonrelativistic matter domination era $T \leq1$eV. This is far smaller than
the minimum bound obtained for $m_{\phi}$ especially in our model in  which the scalar field is coupled to matter and so the mass of the  scalar field is dependent  on local matter. Hence, in the nonrelativistic phases of matter
evolution the decay rate is simply$ \Gamma = \alpha_{\phi} m_{\phi}$.\\
In order to obtain better insight, we need to know
the numerical values of the model parameters. As mentioned above,
 the order of magnitude of $\alpha_{\phi} m_{\phi}$ is about the order of magnitude of
the decay width, which is the reciprocal of the characteristic time scale of reheating. The
inflationary era ends, and reheating can start at the earliest at around $t = 10^{-32}$ s, while
the hot big bang commences at around $t = 10^{-18}$ s \cite{18, 19}. The reheating process should be
complete before the hot big bang to restore the big bang Nuvleosynthesis. Therefore $10^{18} {\rm s^{-1}} \leq \Gamma \leq 10^{32}{\rm s}^{-1}$ and this requires that
\begin{equation}\l{57}
1 {\rm KeV}\leq \alpha_{\phi} m_{\phi}\leq 10^8 {\rm GeV}.
\end{equation}
Note that in our model $m^2_{\phi}$ is time dependent and this fact was one of the main insights of inflationary and reheating cosmology
 in the  1990s.  to explain the preheating we must couple the inflaton field $\varphi$ to another matter scalar field through an interaction term in the Lagrangian (\ref{1}) as
\begin{equation}\l{58}
{\cal L}_{int}= (1/2)g^2\phi^2\psi^2,
\end{equation}
 where g is a
dimensionless coupling constant.
The total effective potential for this system will
be the sum of the  effective potential, $V_{eff}$ driving inflation which was independent of $\psi$ and  the
above interaction term:
\begin{equation}\l{59}
U_{eff} = V_{eff}(\phi) + (1/2)g^2\varphi^2\psi^2.
\end{equation}
Note that the inflaton field has two interactions with matter in this model. One of that is created with  the chameleon mechanism and the other one is created as (\ref{58}). According to the above equation, $\psi$, with zero bare mass, will find an effective mass as
\begin{equation}\l{60}
m_{\psi} = g \phi(t).
\end{equation}
It is well-known, in our model, the  $m_{\psi}$ is much smaller than $m_{\phi}$ and so according to [87, 237] this plys the crucial role in
 the reheating process.
So one can obtain the Fourier modes of the $\psi$ field which obey the following relation
\begin{equation}\l{61}
\ddot{\psi}_k + 3H\dot{\psi}_k = -{\bigr [}{k^2 \over a^2} + g^2\phi^2(t){\bigl ]} \psi_k,
\end{equation}
Using $\psi = a^{-3/2}X(t)$,  where $a(t)  $  is the scale of the  FLRW Universe, we have
\begin{equation}\l{62}
\ddot{X}(t) -\omega_k^2X(t) =0,\end{equation}
where
\begin{equation}\l{63}
\omega_k = \sqrt{{k^2\over a^2} -{3\over2}\dot{H}- {9\over 4} H^2 + g^2\phi(t)^2}.
\end{equation}
It is seen that this equation is an oscillating equation with time dependence frequency $\omega_k$  and the crucial parameter in this $\omega$ is $m^2_X$. If this quantity, $m^2_X$ is changing rapidly   then $\omega_k$ change also. This is quantified by the dimensionless parameter $R_a$ which is defined as
\be\l{64}
R_a \equiv {\dot{\omega}_k \over \omega^2_k}.
\ee
The $R_a \ll 1 $ is usually known as the adiabatic regime. In this regime there is no particle creation and so the number of particles is constant. But if  $R_a \gg 1 $ the number of particles is not constant; this means that in this regime of the  model there is particle creation during the reheating.
One can obtain the adiabaticity parameter as
\be\l{641}
{|\dot{\omega}_k| \over \omega^2_k} \sim {|\dot{\varphi}| \over g\varphi^2}.
\ee
In the interval $|\Delta\varphi| \leq \sqrt{|\dot{\varphi}| / g}$  the adiabaticity parameter, Eq. (\ref{641}),  is greater than  ${\cal O}(1)$. Here $\dot{\varphi}$ is evaluated at the collision time.\\
To obtain an explicit relation for the  reheating temperature, e-folding number and other relevant quantity  the reheating process needs further study and investigation  for some typical examples. So this still is an open problem in our model.}
\subsection{ Late time}
The present model is a slow-rolling model. The Percentage change in the expectation value of $\phi$ in one Hubble time, $\tau = \dot{\phi}/\phi H $ is a criterion for the slow-rolling limit of a scalar field. If this quantity be $\tau \approx V_{eff, \phi}/\phi H$; so, if this ratio is sufficiently small, the scalar field is in the slow-rolling limit. In this case $\dot{\phi}$  is very small and we can illegal $\dot{\phi}^2$.
 Therefore we arrive at
\begin{eqnarray}\label {65}
\frac{\ddot{a}}{a} & \simeq & -\frac{\kappa_5^2}{3} \left(
\frac{1}{2M_5^3} (2+3\omega) +
\frac{\beta^2}{4M_p^3}(1-3\omega)^2 \right) \nonumber \\
 & & \times \left( \frac{\tilde{\rho}_0}{a_0^{3(1+\omega)}} \right)^2\exp{(\frac{2(1-3\omega)\beta\phi}{M_p^{3/2}})}.
\end{eqnarray}
If we take the magnitude of $M_5$ in order of electroweak scale
$M_E=1TeV$ {\cite{11-c}} and $M_p$ in order of $10^{18}GeV$, the
second term in parenthesis may be ignored with respect to the
first term. So, to have an accelerated expanding Universe in this
area $\omega$ should obey as follows
\begin{equation}\label {66}
\omega<-\frac{2}{3},
\end{equation}
this range for $\omega$ is consistent with astronomical data,
because the most recent data indicate that $\omega<-0.76$ at the
$95\%$ confidence level{\cite{11-d}}.  This consistency  shows that in this epoch the two models
   ( standard cosmology and brane world cosmology with chameleon scalar field in the bulk) have the same observational results, because
        in this epoch( late time ) one can ignore $\rho^2$ term in Eq. (\ref{35}) and then this  model can be reduces to the standard cosmology model.  For
$-1<\omega<-\frac{2}{3}$, with attention to the power of $a_0$,
there is a positive and decreasing value for $\ddot{a}$ and for
$\omega<-1$ there is a positive and increasing value.

\section{Validity of Generalized Second Law of Thermodynamics}
{ One important question concerns the thermodynamics behavior of the  Universe. The first connection between general relativity and thermodynamics was given by Bekenstein in 1973. Since then, thermodynamics aspect of the  Universe has been the subject of several studies and  the validity of the  thermodynamics law has  been investigated in many works. In recent years, therehas been  a lot of interest in the brane world scenario which  has given us a new, interesting picture of the  Universe and a generalized form of the  Universe. Maybe the main reason for studying the validity of the  thermodynamics law is that it is natural to study this subject for models which have been built in this new scenario.} \\
Let us now examine the validity of the GSLT, with the  assumption of  the validity the first law.
We consider a region  of a FLRW Universe involved with the  horizon. There
are some works for the  validity of the  GSLT in the Dvali–Gabadadze–Porrati (DGP) brane world scenario
{\cite{12,13}}. This bounded region is filled with a perfect
fluid, where
$$p_b=\omega \rho_b.$$ The  amount of energy crossing the horizon
in time $dt$ is equal to
\begin{equation}\label {67}
\frac{dE}{dt}=\frac{4\pi}{3}R_h^3 \left( -3H(\rho_b+p_b) +
2\phi'\dot{\phi} \right),
\end{equation}
where $R_h$ is the radius of the horizon. With the help of the
validity of the first law of thermodynamics, variation of horizon
entropy is expressed as
\begin{equation}\label {68}
\dot{S}_h=\frac{4\pi R^3_h}{3T_h} \left(
3H(\rho_b+p_b)-2\phi'\dot{\phi} \right)
\end{equation}
$T_h$ is the temperature of the  horizon. Using the Gibbs's
equation,$$T_h dS_I=dE_I + p_b dV.$$ Note that according to
{\cite{14}}, we have supposed an equilibrium for the  temperature
inside matter and horizon. Time evolution of the entropy  inside the
horizon is obtained as
\begin{equation}\label {69}
\dot{S}_I=\frac{1}{T_h} \left( V\dot{\rho}_b+(\rho_b+p_b)\dot{V}
\right).
\end{equation}
Since we have $V=\frac{4\pi}{3}R_h^3$ and $E=\rho_b V$,
$\dot{S}_I$ can be rewritten as
\begin{eqnarray}\label {70}
\dot{S}_I &=&\frac{4\pi R^2_I}{T_h} \nonumber \\
 &\times &\left( \frac{R_I}{3}(-3H(\rho_b+p_b)+2\phi'
\dot{\phi})+(\rho_b+p_b)\dot{R}_I \right).
\end{eqnarray}
Adding Eqs.(\ref{68}) and (\ref{70}), we attain the total
variation of entropy
\begin{eqnarray}\label {71}
\dot{S}&=&\dot{S}_h+\dot{S}_I=\frac{4\pi
R_h^2}{T_h}(\rho_b+p_b)\dot{R}_I
\end{eqnarray}
Now, we are ready to examine the validity of the  GSLT for the  apparent
horizon and cosmological event horizon. In the following we
compute the time variation of $R$ for both of these horizons.
\begin{itemize}

\item{\bf Apparent horizon:}\\
For the  flat space of our  Universe, the  apparent horizon is described by
$R_A=\frac{1}{H}$. So we have
\begin{equation}\label {72}
\dot{R}_A=-\frac{\dot{H}}{H^2}.
\end{equation}
The sign of $\dot{R}_A$ depends on the sign of $\dot{H}$.
\item{\bf Cosmological event horizon:}\\
The radius of the cosmological event horizon is given by
$$R_E=a_0(t)\int_t^\infty \frac{dt}{a_0(t)}.$$ So the time
variation of $R_E$ is described by
\begin{equation}\label {73}
\dot{R}_E=HR_E-1.
\end{equation}
According to {\cite{15}}, for a Universe with a positive
acceleration, like our model, if $\dot{H}>0$ , there is
\begin{equation}\label {74}
\frac{\ddot{a}}{a} - \frac{\dot{a}^2}{a^2} >0
\end{equation}
by integration above relation
\begin{equation}\label {75}
\int_t^\infty \frac{d\dot{a}}{\dot{a}^2}>\int_t^\infty
\frac{dt}{a}.
\end{equation}
For an accelerated expanding Universe, $\ddot{a}>0$, in the late
time $\dot{a}(t\longrightarrow\infty)$ goes to infinty. We can
compute the above integral as
\begin{equation}\label {76}
\dot{a}\int_t^\infty \frac{dt}{a}<1,
\end{equation}
which  means $HR_E<1$. Therefore there is $\dot{R}_E<0$ for
$\dot{H}>0$ and $\dot{R}_E>0$ for $\dot{H}<0$.
\end{itemize}
With attention to the above result, one sees that for investigation of
the validity of GSLT we need to determine the sign of $\dot{H}$.
We carry it out  with the help of  the first Friedmann equation. By
taking the  time derivative of the first Friedmann equation, we arrive
at
\begin{eqnarray}\label {77}
2H\dot{H} & =& -\frac{\kappa_5^4}{6}H(1+\omega)\rho_b^2+\frac{\kappa_5^2}{9}\rho_b
\phi'\dot{\phi} \nonumber \\
 & &+ \frac{\kappa_5^2}{3}\dot{\phi}
(\ddot{\phi}+V_{,\phi}(\phi))  +\frac{\kappa_5^2}{3}\phi'\dot{\phi}'.
\end{eqnarray}
The reader should note some points: first, we have an
accelerated expanding Universe, so $H$ is always positive;
second, if we suppose that the scalar field is  a uniform
decreasing function of time, then the time derivative of it is
negative; third, $V_{,\phi}(\phi)$, as we can see in {\cite{5,6}},
is always negative, so $\dot{\phi}V_{,\phi}>0.$ Substituting the
energy density and with the help of the  junction condition, one can
rearrange the above equation
\begin{eqnarray}\label {78}
2H\dot{H}&=&-\frac{\kappa_5^2}{3} \tilde{\rho}_0^2
\exp{(\frac{2(1-3\omega)\beta\phi}{\sqrt{M_p^3}})}
a^{-6(1+\omega)} \nonumber \\
 & &\Bigg\{ \kappa_5^2 \left(
-\frac{H(1+\omega)}{2} +
\frac{\beta}{3\sqrt{M_p^3}}(1-3\omega)\dot{\phi} \right) \nonumber \\
 & &+\frac{\beta^2 (1-3\omega)^2}{M_p^3}\left(
\frac{(1-3\omega)\beta\phi}
 {\sqrt{M_p^3}} - H \right)
\Bigg\} \nonumber \\
 & & + \frac{\kappa_5^2}{3}\dot{\phi}\left( \ddot{\phi} +
V_{,\phi}(\phi) \right)
\end{eqnarray}
Since the $\dot{\phi}$ is small and because of the existence of
the $M_p$ in the denominator we can estimate
\begin{eqnarray}\label {79}
2H\dot{H} & \longrightarrow & -\frac{\kappa_5^2}{6}
H(1+\omega)\tilde{\rho}_0^2
\exp{(\frac{2(1-3\omega)\beta\phi}{\sqrt{M_p^3}})}
a^{-6(1+\omega)} \nonumber \\
 & & + \frac{\kappa_5^2}{3} \dot{\phi} V_{,\phi}(\phi)
\end{eqnarray}
to have $\dot{H}>0$; especially in early times, there should be
$\omega<-1$. This result indicates that $\dot{H}>0$, so the time
variation of both the  apparent horizon and cosmological event horizon
is  negative, namely $\dot{R}_E,\dot{R}_A<0$. For validity  of the
GLST the equation of state parameter should be smaller than $-1$;
this is compatible with relation (\ref{26}). In this range of
$\omega$ our matter is phantom. If $-1<\omega<-\frac{2}{3}$,
there is still an accelerated expansion, and for validity of GSLT,
$\dot{H}$ should be negative. It means the first term on the
right-hand side of (\ref{44}) should be dominant. With considering
equation (\ref{44}) we see that this condition is not
unreasonable. In this range value of $\omega$ our matter is named
quintessence.

\section{Conclusion}
In this paper a kind of scalar field is embedded in the bulk
space-time, called the  chameleon scalar field. { A brief review on gravity localization has been mentioned, whereby taking some gauge as well as the  transverse-traceless gauge, a wave equation for the perturbation of the metric could be achieved to let us determine whether or not gravity localized on the  brane.} The junction conditions
are  acquired. The junction condition for the  component of the  metric is
similar to the result of the work of Binetruy $\textit{et al}$, but
in contrast to their result, here the components of the
energy-momentum tensor depend on the scalar field with an
exponential term. Considering the generalized Friedmann equation
tells us there is an exponential expansion in early times. This
exponential expansion may have relation with the inflation period. { The assumption of interaction between matter and the  scalar field brings us an effective potential to recover the  basic equation for explaining inflation. It has been shown that in the  radiation dominant era, the exponential term is omitted from the  acceleration equation.}
Also we have a positive acceleration in late time for the  Universe.
The equation of motion of the  scalar field on the brane is similar to
the standard model and it displays that the mass of the scalar field
depends on the local energy density. Investigation of the  validity
of the  generalized second law of thermodynamics for the apparent horizon
and cosmological event horizon shows that the matter on the brane
can be phantom or quintessence.

\section{Aknowledgement}
The authors thank the referee for his/her illuminating
remarks that enabled them to improve the clarity of the paper. The work of  Kh. Saaidi has been supported financially  by  the University of Kurdistan, Sanandaj, Iran, and  he would like thank to the University of Kurdistan for supporting him in his sabbatical period.



\begin{thebibliography}{99}

\bibitem{1} S. Perlmutter et al., Nature {\bf{391}}, 51, (1998); Astrophys. J.
{\bf{517}}, 565, (1999).

\bibitem{2} R. R. Caldwell, R. Dave and P. J. Steinhardt, Phys. Rev. Lett. {\bf{80}}, 1582 (1998); P.
J. E. Peebles and A. Vilenkin, Phys. Rev. D {\bf{59}}, 063505
(1999); P. J. Steinhardt, L. M. Wang and I. Zlatev, Phys. Rev. D
{\bf{59}}, 123504 (1999); M. Doran and J. Jackel, Phys. Rev. D
{\bf{66}}, 043519 (2002); A. R. Liddle, P. Parsons and J. D.
Barrow, Phys. Rev. D {\bf{50}}, 7222 (1994).

\bibitem{3} R. R. Caldwell, Phys. Lett. B {\bf{545}}, 23 (2002); R. R. Caldwell, M.
Kamionkowski and N. N. Weinberg, Phys. Rev. Lett. {\bf{91}},
071301 (2003); J. M. Cline, S. Y. Jeon and G. D. Moore, Phys.
Rev. D {\bf{70}}, 043543 (2004).

\bibitem{4} G. R. Dvali, G. Gabadadze and M. Porrati, Phys. Lett. B {\bf{485}}, 208 (2000); G. R.
Dvali, G. Gabadadze, M. Kolanovic and F. Nitti, Phys. Rev. D
{\bf{64}} 084004 (2001).

\bibitem{5} J. Khoury, A. Weltman, Phys. Rev. D {\bf{69}}:044026,
(2004).

\bibitem{6} T. P. Waterhouse, [arXiv:astro-ph/0611816].

\bibitem{7} P. Brax, C. van de Bruck, A.-C. Davis, JCAP {\bf{0411}}, 004, (2004).

\bibitem{8} L. Randall and R. Sundrum, Phys. Rev. Lett. {\bf{83}}, 3370, (1999); Phys. Rev. Lett.
{\bf{83}}, 4690, (1999).
\bibitem{8'} E. Abou El Dahab, S. Khalil, JHEP, {\bf 0609}; 042 , (2006).
\bibitem{9} T. Shiromizu, K. I.  Maeda and M. Sasaki, Phys. Rev. D
{\bf{62}}, 024012, (2000).

\bibitem{10} P. Binetruy, C. Deffayet, D. Langlois, Nucl. Phys. B {\bf{565}}:269-287, (2000); P. Binetruy,
C. Deffayet, U. Ellwanger, D. Langlois, Phys. Lett. B {\bf{477}}:285-291, (2000).

\bibitem{10-a} D. Langlois, M. Rodriguez-Martinez, Phys. Rev. D \\
{\bf 64},123507, (2001).

\bibitem{11} H. A. Chamblin and H. S. Reall, Nucl. Phys. B {\bf{562}}, 133, (1999).

\bibitem{11-a} C. Bogdanos, A. Dimitriadis, K. Tamvakis, Class. Quant. Grav. {\bf{24}}:3701-3712,
(2007); D. Langlois, M. Rodriguez-Martinez, Phys. Rev. D
{\bf{64}}, 123507, (2001); S. C. Davis, JHEP {\bf{0203}}, 054, (2002).

\bibitem{11-b} K. Saaidi, A. Mohammadi, Modern Phys. Lett. A {\bf{25}}, 3061-3068 (2010).
\bibitem{11-c} P. Brax, C. van de Bruck, Class. Quant. Grav. {\bf{20}}: R201-R232, (2003).

\bibitem{11-d} A. G. Riess, et al., Astrophys. J. {\bf{607}}:665-687, (2004); M. Li, Phys. Lett. B {\bf{603}}, 1, (2004).

\bibitem{12} J. Dutta, S. Chakraborty, M. Ansari, Mod. Phys. Lett. A {\bf{25}}:3069-3079,
(2010); J. Dutta, S. Chakraborty, Gen. Relativity
Gravit. {\bf 42}, 1863, (2010).

\bibitem{13} A. Sheykhi, B. Wang, R.-G. Cai, Nucl. Phys. B {\bf{779}}:1-12,
(2007); R.-G. Cai, L.-M. Cao, Nucl. Phys. B {\bf{785}}:135-148, (2007).

\bibitem{14} E. N. Saridakis, P. F. Gonzalez-Diaz, C. L. Siguenza,
Class. Quant. Grav. 26:165003, (2009); Nucl. Phys. B {\bf{697}},
363, (2004).

\bibitem{15} H. Mohseni Sadjadi, Phys.Rev. D {\bf{73}}, 063525, (2006).
\bibitem{16} A. H. Guth, Phys. Rev. D{\bf 71}, 347 (1981).
\bibitem{17} A. Linde, Phys. Repts. {bf 575},  333, (2000); B. A. Bassett, S. Tsujikawa and D. Wands, Rev.
Mod. Phys. {\bf 78}, 537 (2006).
\bibitem{18} E. W. Kolb and M. S. Turner, The Early Universe (Addison-Wesley, Menlo Park, CA) (1990).
\bibitem{19} C. D. Hoyle et al., Phys. Rev. Lett. {\bf  86}, 1418 (2001).
\bibitem{20}A. D. Dolgov and A. D. Linde, Phys. Lett. B{\bf 116}, 329 (1982); L. F. Abbott, E. Farhi and M.
B. Wise, Phys. Lett. B{\bf 117}, 29 (1982); A. Albrecht, P. J. Steinhardt, M. S. Turner and F.
Wilczek, Phys. Rev. Lett. {\bf 48}, 1437 (1982).
\bibitem{21}T. Harko and M. K. Mak, Astrophys. and Space Science, {\bf 253}, 161 (1997).
\bibitem{22} C. Csaki, J. Erlich, T. Hollowood and Y. Shirman,   Nucl. Phys. B {\bf 581} 309 (2000); A. Brandhuber and K.  Sfetsos  JHEP {\ref 9910},  013 (1999).

\bibitem{24} T. Gherghetta, \textit{et.al.}, Phys. Rev. Lett. {\bf 85}, 240-243 (2000); A. Herrera-Aguilar \textit{et. al.}, JHEP  {\bf 1011}:015, (2010); D. Langlois, Phys. Rev. D {\bf 62}, 126012 (2000); M. ITO, Europhys. Lett. {\bf 64}:295-301, (2003).

\bibitem{25}O.  DeWolfe, D. Z. Freedman, S. S. Gubser and A.  Karch,  Phys. Rev. D {\bf 62} 046008 (2000).

\bibitem{26} Heng Guo, \textit{et. al.}, [arXiv:1008.3686v2].


\bibitem{27} S. V. Chervon, Gen. Rel. and Grav. 36 1547 (2004); V. M. Zhuravlev and S. V. Chervon, Zh. Exp. Teor. Fiz. 118, 259 (2000); S. V.
Chervon and V. M. Zhuravlev, Nucl. Phys. B 80 (Proc. Suppl.) (2000).

\bibitem{28} B.C. Paul∗and Dilip Paul, [arXiv:0708.0897v1].

\bibitem{29} R. Maartens, D. Wands, B. A. Bassett, I. P. C. Heard, Phys. Rev. D {\bf 62}, 041301 (2000).

\bibitem{30} Edmund J. Copeland, Andrew R. Liddle, James E. Lidsey, [arXiv:astro-ph/0006421v2]
\bibitem{31}A. D. Dolgov and A. D. Linde, Phys. Lett. B {\bf 116}, 329 (1982).
\bibitem{32} L. F. Abbott, E. Fahri and M. Wise Phys. Lett. B {\bf 117},
29 (1982).
\bibitem{33} E. W. Kolb, A. Notari and A. Riotto, Phys. Rev. D {\bf 68}, 123505 (2003).
\bibitem{34}E. W. Kolb and M. S. Turner, The Early Universe (Addison-Wesley, Menlo Park, CA) (1990).
\end{thebibliography}
\end{document}